\documentclass[prl,twocolumn,unsortedaddress,superscriptaddress]{revtex4-1}
\usepackage[T1]{fontenc}
\usepackage[utf8]{inputenc}
\usepackage{graphicx}
\usepackage{amsmath}
\usepackage{color}
\usepackage{comment}
\usepackage{hyperref}
\usepackage{multirow}
\usepackage{subfig}
\usepackage[left=1.5cm,right=1.5cm,top=2cm,bottom=2cm]{geometry}

\begin{document}

\title{Full configuration interaction quantum Monte Carlo benchmark and multireference coupled cluster studies of tetramethyleneethane diradical}

\author{Libor Veis}
\email{libor.veis@jh-inst.cas.cz}
\affiliation{J. Heyrovsk\'{y} Institute of Physical Chemistry, Academy of Sciences of the Czech \mbox{Republic, v.v.i.}, Dolej\v{s}kova 3, 18223 Prague 8, Czech Republic}

\author{Andrej Antal\'{i}k}
\email{andrej.antalik@jh-inst.cas.cz}
\affiliation{J. Heyrovsk\'{y} Institute of Physical Chemistry, Academy of Sciences of the Czech \mbox{Republic, v.v.i.}, Dolej\v{s}kova 3, 18223 Prague 8, Czech Republic}

\author{\"Ors Legeza}
\email{legeza.ors@wigner.mta.hu}
\affiliation{Strongly Correlated Systems ``Lend\"{u}let'' Research group, Wigner Research Centre for Physics, H-1525, Budapest, Hungary}

\author{Ali Alavi}
\email{A.Alavi@fkf.mpg.de}
\affiliation{Department of Chemistry, University of Cambridge, Cambridge, United Kingdom}
\affiliation{Max Planck Instit\"{u}t f\"{u}r Festk\"{o}rperforschung, Stuttgart, Germany}

\author{Ji\v{r}\'{i} Pittner}
\email{jiri.pittner@jh-inst.cas.cz}
\affiliation{J. Heyrovsk\'{y} Institute of Physical Chemistry, Academy of Sciences of the Czech \mbox{Republic, v.v.i.}, Dolej\v{s}kova 3, 18223 Prague 8, Czech Republic}

\date{\today}

\begin{abstract}
We have performed a FCI-quality benchmark calculation for the tetramethyleneethane molecule in cc-pVTZ basis set employing
a subset of CASPT2(6,6) natural orbitals for the FCIQMC calculation.
The results are in
an excellent agreement with the previous large scale diffusion Monte Carlo calculations by Pozun \textit{et al.} and available experimental results. 
Our computations verified that there is a maximum on PES of the ground singlet state ($^1\text{A}$) $45^{\circ}$ torsional angle and the corresponding vertical singlet-triplet energy gap is $0.01$ eV.
We have employed this benchmark for assessment of the accuracy of MkCCSDT and DMRG-tailored CCSD (TCCSD) methods.
MR MkCCSDT with CAS(2,2) model space, though giving good values for the singlet-triplet energy gap, is not able to properly describe the shape of the multireference singlet PES. 
Similarly, DMRG(24,25) is not able to correctly capture the shape of the singlet surface, due to the missing dynamic correlation.
On the other hand, the DMRG-tailored CCSD method describes the shape of the ground singlet state with an excellent accuracy, but for the correct ordering requires computation of the zero-spin-projection component of the triplet state ($^3\text{B}_1$).

\end{abstract}


\maketitle

\section{Introduction}
\label{section_introduction}

\begin{figure}[!ht]
  \begin{center}
    \hskip -0.5cm
    \includegraphics[width=0.45\textwidth]{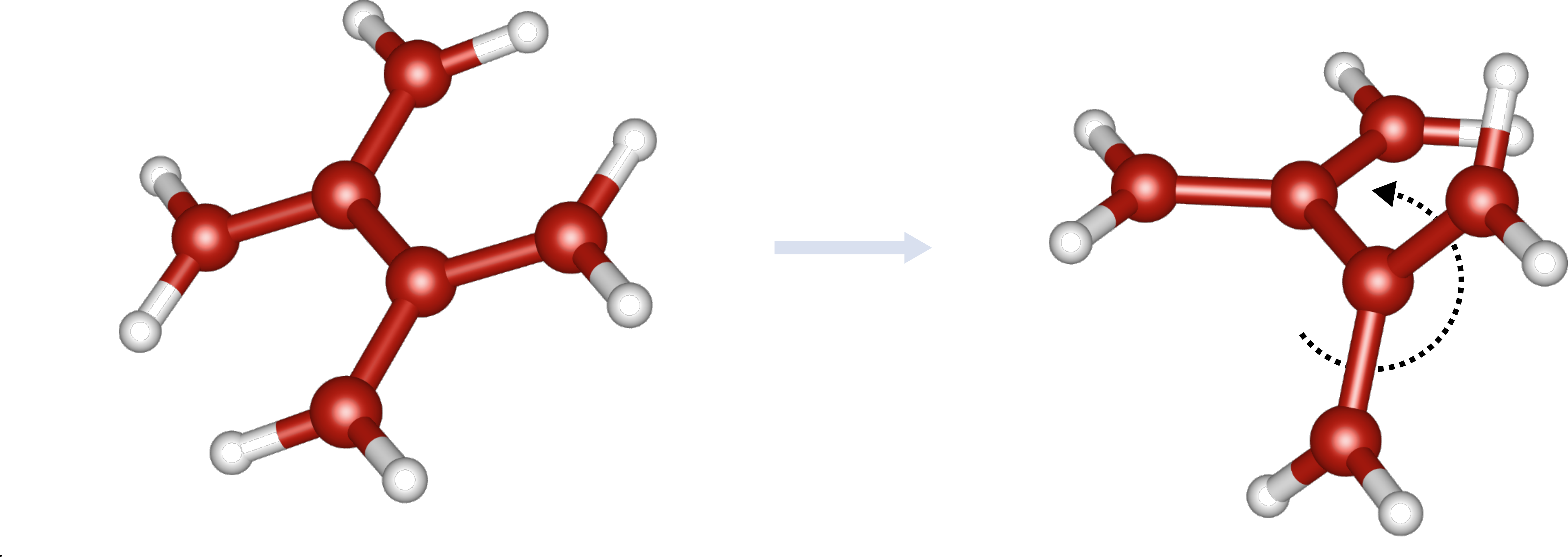}
    \caption{The studied process of a rotation of the TME allyl subunits about the central C-C bond. Carbon atoms are colored brown, hydrogens are white.}
    \label{tme_rotation}
  \end{center}
\end{figure}

Tetramethyleneethane (TME), the simplest disjoint non-Kekul\'{e} diradical firstly synthesized by Dowd \cite{dowd_1970}, has due to its complex electronic structure been often used as a benchmark system for the state-of-the-art multireference computational methods \cite{pittner_2001, bhaskaran-nair_2011, chattopadhyay_2011, pozun_2013, demel_2015}. Its complexity comes out of the fact that it contains a nearly degenerate pair of the frontier orbitals, which tend to be localized on separate allyl subunits \cite{pozun_2013} and  are occupied by two electrons. Moreover, TME possesses a degree of freedom corresponding to the rotation about the central C-C bond (maintaining $D_2$ symmetry, see Figure \ref{tme_rotation}) and the energetic ordering of these two frontier orbitals and consequently their occupation in the lowest singlet state changes along the rotation \footnote{At the torsional angle matching 45$^{\circ}$, the occupation of both frontier orbitals is approaching one.}. As a result, determining the relative stability of the lowest singlet and triplet states turned out to be a big challenge for both experimental and theoretical methods.

The first experimental electron paramagnetic resonance (EPR) results predicted TME to have a triplet ground state \cite{dowd_1970}, when stabilized in a matrix with a torsional angle being approximately 45$^{\circ}$ \cite{dowd_1986, dowd_1987}. The predicted triplet ground state attracted a lot of interest in TME for its potential use as an organic magnet \cite{berson_1999}. However, photo-electron spectroscopy of the TME$^{-}$ ion strongly suggested TME to have the singlet ground state at the torsional angle corresponding to 90$^{\circ}$ \cite{clifford_1998}, similarly as the EPR experiments on TME derivatives \cite{bush_1997a, bush_1997b}.

Several theoretical studies using different level of approximations \cite{du_1987, nachtigall_1992, nachtigall_1993, filatov_1999, rodriguez_2000, pittner_2001, bhaskaran-nair_2011, chattopadhyay_2011} have step by step contributed to understanding of the electronic structure of the TME diradical. Nevertheless, only the work of Pozun \textit{et al.} employing the large scale diffusion Monte Carlo (DMC) calculations \cite{pozun_2013} finally reliably established the magnitude of the singlet triplet gap and also the shape of the singlet potential energy surface (PES).

The main conclusions tell us that the correct theoretical description of the multireference singlet state ($^{1}\text{A}$) requires all of the following conditions being fulfilled, namely the flexible-enough atomic basis set, the proper description of the static correlation with the minimum active space comprising of six $\pi$-orbitals, and a proper treatment of the dynamic correlation (at least at the level of the second-order perturbation theory). All this make TME a very delicate molecule and indeed the perfect benchmark system for state-of-the-art multireference methods. Moreover, TME serves as a model system for more complicated disjoint diradicals.

In the present work, we follow \cite{pozun_2013} and compute the singlet, as well as triplet twisting PESs of TME. Firstly, we provide the full configuration interaction (FCI) quality data by the FCI quantum Monte Carlo (FCIQMC) method \cite{booth_2009, cleland_2010, petruzielo_2012, overy_2015}, whose accuracy is justified by an excellent agreement with DMC results of Pozun \textit{et al.} \cite{pozun_2013} and available experimental data. Secondly, we compare the results of the Hilbert space Mukherjee's multireference coupled clusters (MR MkCC) \cite{mahapatra_1999, evangelista_2006, evangelista_2007, evangelista_2008, evangelista_2010, das_2010, bhaskaran-nair_2008, bhaskaran-nair_2010, demel_2010} and the recently developed coupled clusters with single and double excitations tailored by the density matrix renormalization group method (DMRG-TCCSD) \cite{veis_2016} against the FCIQMC benchmark.

The paper is organized as follows: 
in Sec. II, we give a very brief overview of the used computational approaches and the actual computational details, the next Section summarizes the results with discussion, and the final Section closes with conclusions and outlook.

\section{Overview of computational approaches}
\label{section_computational_details}

In this Section we, for completeness, sketch the main concepts and ideas of the employed computational approaches.

\subsection{FCI quantum Monte Carlo}

The FCIQMC method \cite{booth_2009, cleland_2010, booth_2014, overy_2015}, originally developed by one of us, is a stochastic approach performing a long time integration of the imaginary-time Schr\"odinger equation which is capable of converging onto the FCI solution for much larger orbital spaces than the exact diagonalization allows. In contrast to DMC, FCIQMC sample the \textit{Slater determinant space} by an ensemble of walkers that move around randomly.

Master equations governing walkers’ population dynamics are given by

\begin{equation}
  -\frac{\text{d} N_i}{\text{d} \tau} = (H_{ii} - S)N_i + \sum_{j \neq i} H_{ij} N_j,
  \label{qmc_dynamics}
\end{equation}

\noindent
where $\tau$ is imaginary time, $N_i$ the walker population on determinant $i$, $S$ the energy shift parameter controlling the total walker population, and $H_{ij}$ Hamiltonian matrix elements in the basis of Slater determinants. When employing the stochastic approach, individual walkers evolve according to a simple set of rules which include spawning, death/cloning and most importantly also annihilation processes \cite{booth_2009}.

We have used the semi-stochastic method with real walker weights \cite{petruzielo_2012}, in which part of the imaginary-time propagation (Eq. \ref{qmc_dynamics}) is performed 
exactly (deterministic space) and the rest stochastically. Such an approach in fact greatly reduces stochastic errors.

\subsection{Mukherjee's coupled clusters}

The MR MkCC approach formulated by Mukherjee \textit{et al.} \cite{mahapatra_1999} and later on developed by others, including one of us \cite{evangelista_2006, evangelista_2007, evangelista_2008, evangelista_2010, das_2010, bhaskaran-nair_2008, bhaskaran-nair_2010, demel_2010, brabec_2012a, brabec_2012, demel_2015}, is a state specific Hilbert-space multireference coupled cluster method. Consequently, the MkCC wave function $|\Psi_{\text{MkCC}}\rangle$ is expressed by means of the Jeziorski-Monkhorst ansatz

\begin{equation}
  |\Psi_{\text{MkCC}}\rangle = \sum_{\mu = 1}^{M} c_{\mu} e^{T(\mu)} |\Phi_{\mu}\rangle.
  \label{jeziorski-monkhorst}
\end{equation}

\noindent
In Eq. \ref{jeziorski-monkhorst},
$|\Phi_{\mu}\rangle$ are the reference functions spanning the model space (in our case complete) and $T(\mu)$ the reference-dependent cluster operators. The $c_{\mu}$ coefficients as well as the desired energy are obtained by diagonalization of the effective Hamiltonian matrix, whose elements read

\begin{equation}
  H_{\mu \nu}^{\text{eff}} = \langle \Phi_{\mu} | e^{-T(\nu)} H e^{T(\nu)} | \Phi_{\nu} \rangle,
\end{equation}

\noindent
with $H$ being the Hamiltonian operator.

The MkCC method is superior to the related Hilbert-space multireference method based on the Brillouin-Wigner CC theory due to its exact size extensivity. 
Though not completely free of problems, the MkCC approach is reliable for small model spaces and indeed the method of choice for electronic structure studies of diradicals.
In the present work, we have employed the MR MkCC methods including single and double (MkCCSD) and single, double, and triple excitations (MkCCSDT).

\subsection{DMRG-based tailored coupled clusters}

The tailored CC (TCC) approach was formulated by Kinoshita \textit{et al.} \cite{kinoshita_2005} and belongs to the class of so called externally corrected CC methods.
The TCC wave function expansion employes the split-amplitude ansatz used previously by Piecuch \textit{et al.} \cite{piecuch_1993, piecuch_1994}

\begin{equation}
  \label{eq:TCC}
  | \Psi_\text{TCC} \rangle = e^{T} | \Phi_\text{0} \rangle =  e^{T_\text{ext}+T_\text{CAS}} | \Phi_\text{0} \rangle =  e^{T_\text{ext}} e^{T_\text{CAS}} | \Phi_\text{0} \rangle,
\end{equation}

\noindent
i.e. the cluster operator is split up into its active space part ($T_\text{CAS}$) and the remaining external part ($T_\text{ext}$). Since $| \Phi_\text{0} \rangle$ is a single-determinant reference wave function, both of the aforementioned cluster operators mutually commute, which keeps the methodology very simple. 

The $T_\text{CAS}$ amplitudes are supposed to be responsible for a proper description of the static correlation. They are computed from the complete active space configuration interaction (CASCI) wave function coefficients and are kept frozen during the CC iterations. Only the $T_\text{ext}$ part, which is responsible for a proper description of the dynamic correlation, is being optimized.

Recently, some of us have developed the DMRG-TCCSD method, i.e. coupled clusters with single and double excitations tailored by matrix product state (MPS) wave functions (wave functions produced by the DMRG algorithm \cite{white_1992, schollwock_2011}) \cite{veis_2016}. This approach replaces CASCI of the original TCC method by DMRG and thus allows employing much larger active spaces. It has indeed proven itself a reliable method suitable for difficult multireference problems requiring larger active spaces \cite{veis_2016}.

\subsection{Computational details}

We have performed constrained geometry optimizations for seven values of the torsional angle along the twisting process. Geometries were optimized for both states ($^1\text{A}$, $^3\text{B}_1$) with the complete active space second order perturbation theory (CASPT2)  as implemented in the MOLPRO package \cite{molpro}. The CASPT2 calculations were carried out using the active space comprising of six $\pi$ orbitals, CAS(6,6), and cc-pVTZ basis \cite{dunning_basis}. Only the first 60 CASPT2(6,6) natural orbitals sorted according to their occupation numbers were kept for the correlation treatment by the FCIQMC, MR MkCC, DMRG, and DMRG-TCCSD methods, the rest was dropped. We have chosen this strategy rather than employing a smaller basis, e.g. 6-31G/6-31G*, which would still be manageable by the massively parallel FCIQMC implementation \cite{booth_2014}, as it was clearly demonstrated in \cite{pozun_2013}, that a triple-$\zeta$ basis with f functions on the C atoms is essential for the proper description of the singlet ($^{1}\text{A}$) PES (see Section with results and discussion for further comments).

For the FCIQMC calculations, we have employed the following computational protocol: (1) equilibration computations with 10 million walkers; (2) generation of FCIQMC natural orbitals \cite{overy_2015} (for faster convergence with the number of walkers) with 50-million-walker computations;
(3) subsequent 100, 500, and 1000-million-walker computations with the FCIQMC natural orbitals. We have used the initiator version of the FCIQMC method as implemented in the NECI program package \cite{neci}. Moreover, to greatly reduce stochastic errors, we have employed the semi-stochastic method with real walker weights \cite{petruzielo_2012} and, in case of the largest 1000-million-walker computations, 50 thousand most populated determinants in the deterministic space.

MR MkCC calculations were performed with the complete model space comprising of the frontier orbitals, CAS(2,2). 

In all production DMRG calculations (those used for generation of the active space CC amplitudes), we have employed the dynamical block state selection (DBSS) procedure \cite{legeza_2003a, legeza_2004} with the truncation error criterion set to $5 \cdot 10^{-6}$, which resulted in bond dimensions varying in the range of $1000-8000$. The orbitals for DMRG active spaces were chosen according to their single-orbital entropies ($S_i$), in particular for CAS(6,6) $S_i > 0.3$, in case of CAS(12,12) $S_i > 0.1$, and for CAS(24,25) $S_i > 0.075$. As usually, the DMRG active space orbitals were split-localized \cite{olivares-amaya_2015}. The Fiedler method \cite{barcza_2011, fertitta_2014} was used for optimization of the orbital ordering and DMRG runs were initialized using the CI-DEAS procedure \cite{legeza_2003b,legeza_review}.

In all DMRG-TCCSD calculations, we have employed the frozen core approximation. Apart from the high spin triplet ($m_s = 1$), for the reasons discussed below, we have also calculated the low spin triplet components ($m_s = 0$). Such calculations were indeed realized by swapping (rotation) of the open-shell $\beta$ spin-orbitals and finally closed-shell computations employing the unrestricted versions of the DMRG \footnote{Huge flexibility of the Budapest QC-DMRG program \cite{budapest_qcdmrg} allows among others use of unrestricted molecular orbital integrals, as well as more general relativistic ones \cite{knecht_2014} or those appearing in nuclear structure calculations \cite{legeza_2015}.} and TCCSD codes (the molecular orbital integrals become spin-dependent).

\section{Results and discussion}
\label{section_results}

\begin{figure}[!ht]
  \begin{center}
    \includegraphics[width=0.45\textwidth]{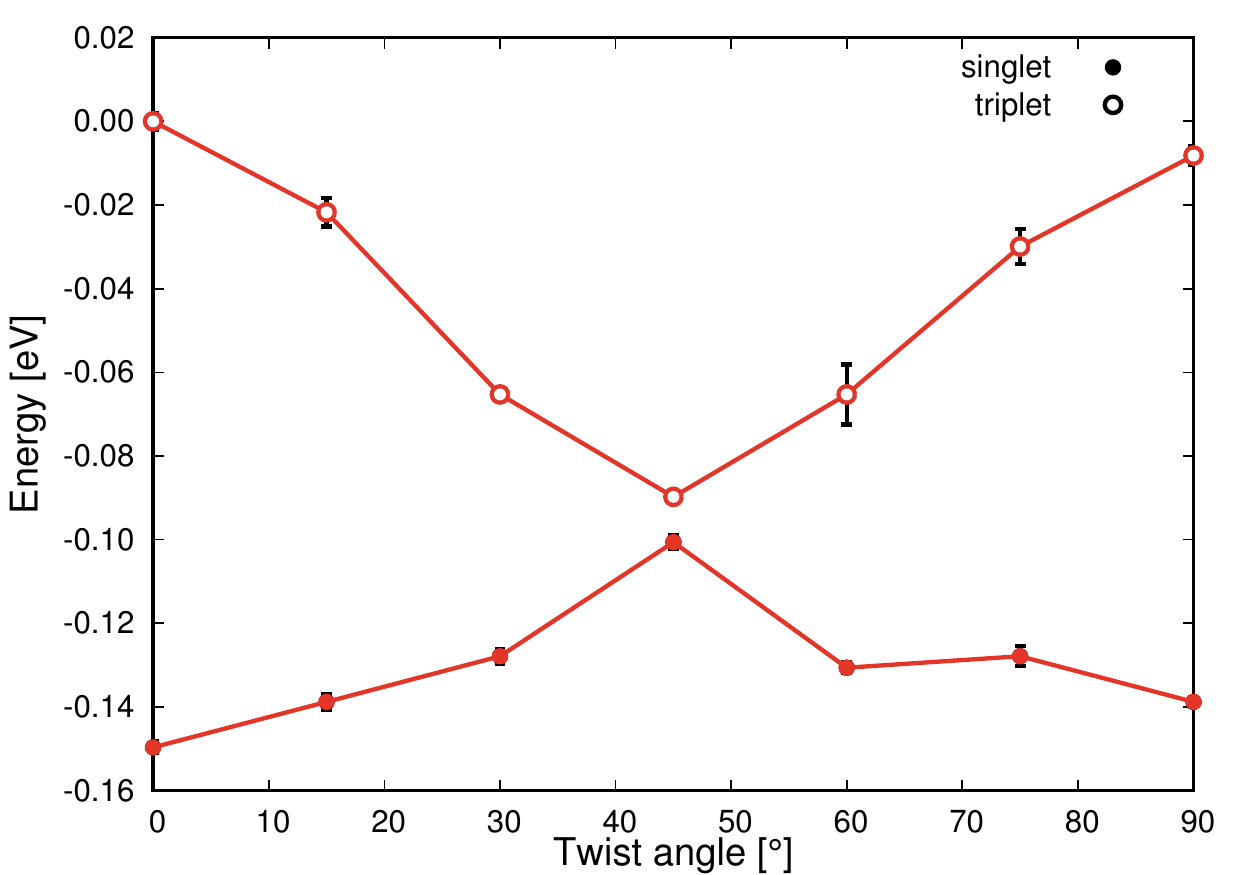} 
    \caption{The FCIQMC singlet ($^1\text{A}$) and triplet state ($^3\text{B}_1$) twisting PESs of TME. Vertical lines correspond to errors calculated by the blocking analysis \cite{flyvbjerg_1989}.}
    \label{fciqmc_plot}
  \end{center}
\end{figure}

The FCIQMC PESs of the singlet ($^1\text{A}$) as well as the triplet state ($^3\text{B}_1$) corresponding to the twisting process are shown in Figure \ref{fciqmc_plot}. We do not present the absolute energies as they may not be fully converged with the number of walkers \footnote{The MR MkCCSDT energies are in fact lower by approximately 10 mHartree, nevertheless the CC method is generally not variational and the error coming out from this fact is questionable.}, however 1000 million walkers was the maximum we could afford with 2000 CPU cores and the relative energies are definitely not affected giving excellent agreement with the DMC energies by Pozun \textit{et al.} \cite{pozun_2013} and available experimental data (see Table \ref{tab_results}).

One can observe a very similar shape of the singlet PES as demonstrated by Pozun \textit{et al.} \cite{pozun_2013}, i.e. with its  maximum corresponding to the torsional angle of 45$^{\circ}$. The height of this ``hump'' [$E(45^{\circ}) - E(0^{\circ})$] calculated by the FCIQMC method equals 0.05 eV. Pozun \textit{et al.} \cite{pozun_2013} demonstrated that a triple-$\zeta$ basis with f functions on the C atoms is essential to obtain a correct shape of the singlet PES with the maxima at 45$^{\circ}$. We have performed additional FCIQMC calculations for the torsional angles of $0^{\circ}$, $45^{\circ}$, and $90^{\circ}$ with the hybrid 6-31G/6-31G* including polarization functions only on the two central C atoms (74 molecular orbitals in total) to verify this conclusion. In fact, the energy difference between the points at $45^{\circ}$ and $0^{\circ}$ that we obtained was zero within the statistical errors, which is in agreement with \cite{pozun_2013}.

The magnitude of the TME singlet-triplet energy gap is indeed very small, corresponding to $0.01$ eV as obtained by the FCIQMC method. This is also the reason for the originally wrong ground state triplet assignment by EPR spectroscopy \cite{dowd_1970}. Weak EPR signal was apparently caused by small population of the triplet state allowed by the rotation about the central C-C bond.

\begin{comment}
\begin{table}[!ht]
  \begin{tabular}{c c c}
    \hline
    Torsional angle & ~~~$^1\text{A}$ state~~~ & ~~~$^3\text{B}_1$ state~~~ \\
    \hline
      0$^{\circ}$ & -232.5492 & -232.5437 \\
    15$^{\circ}$ & -232.5488 & -232.5445 \\
    30$^{\circ}$ & -232.5484 & -232.5461 \\
    45$^{\circ}$ & -232.5474 & -232.5470 \\
    60$^{\circ}$ & -232.5485 & -232.5461 \\
    75$^{\circ}$ & -232.5484 & -232.5448 \\
    90$^{\circ}$ & -232.5488 & -232.5440 \\
    \hline
  \end{tabular}
  \caption{The FCIQMC absolute energies in a.u. calculated with 1000 million walkers.}
  \label{fciqmc_tab}
\end{table}
\end{comment}

\begin{figure}[!ht]
  \begin{center}
    \includegraphics[width=0.45\textwidth]{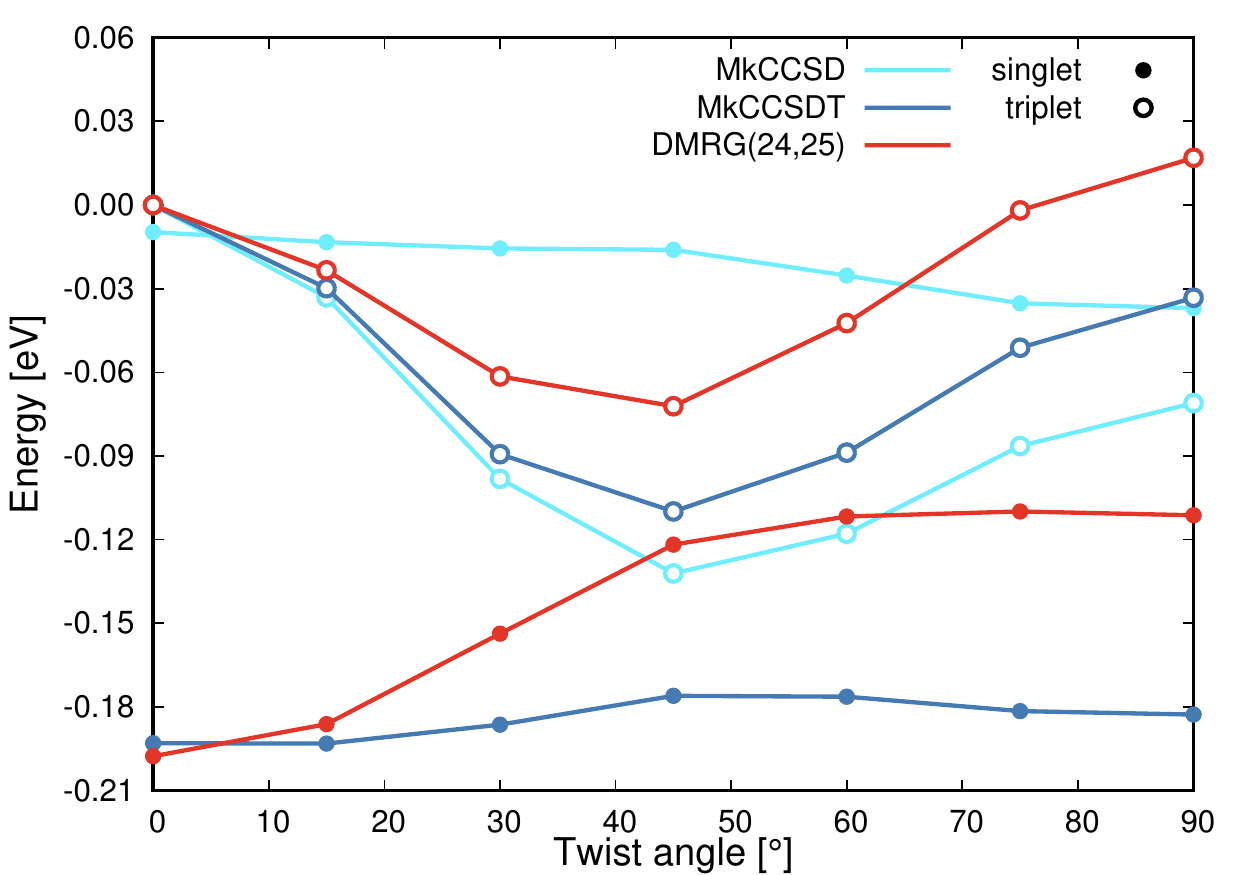} 
    \caption{The MR MkCCSD, MkCCSDT, and DMRG(24,25) singlet ($^1\text{A}$) and triplet state ($^3\text{B}_1$) twisting PESs of TME.}
    \label{mrcc_plot}
  \end{center}
\end{figure}

\begin{figure*}[!ht]
  \subfloat[][]{
    \hskip -0.5cm
    \includegraphics[width=0.45\textwidth]{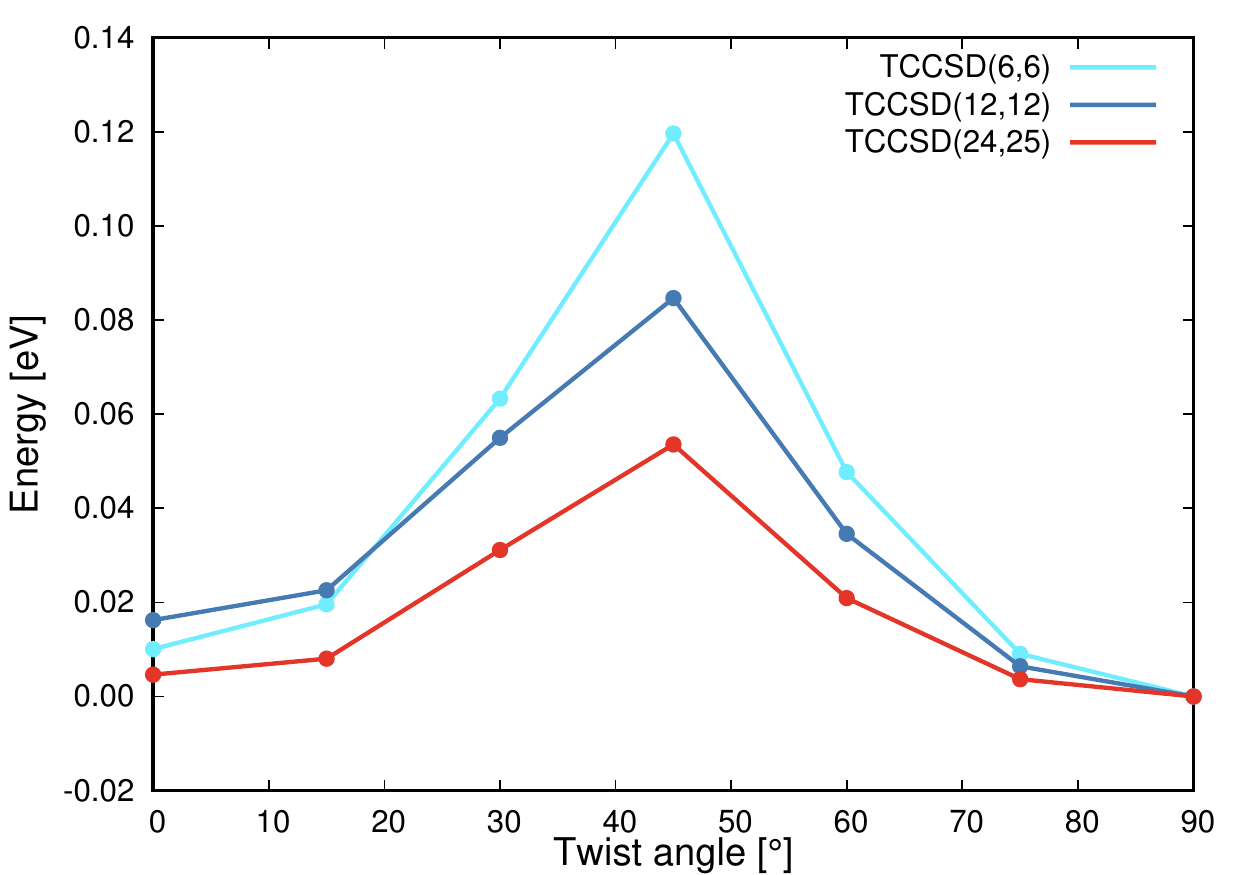} 
    \label{tcc_plot1}
  }
  \subfloat[][]{
    \includegraphics[width=0.45\textwidth]{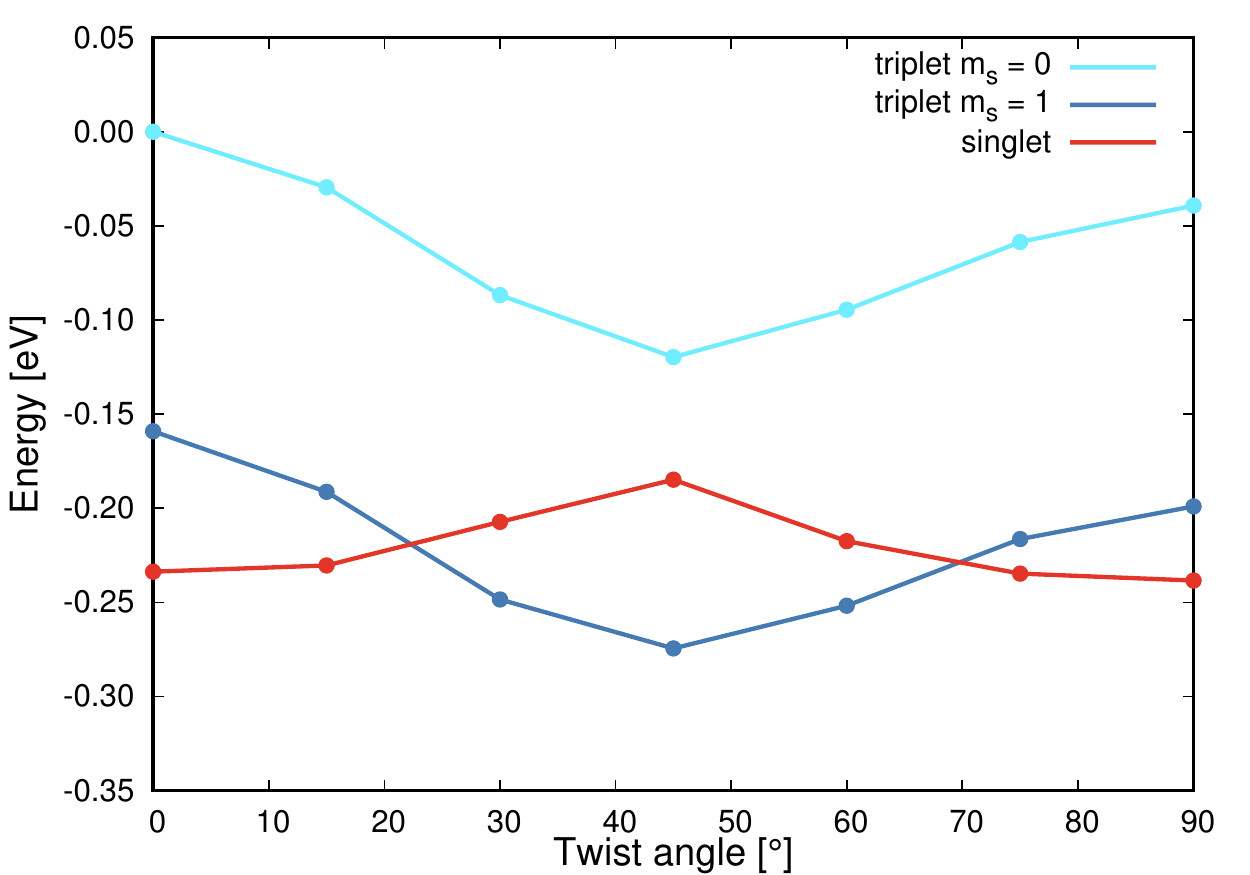} 
    \label{tcc_plot2}
  }
  \caption{The TCCSD PESs of TME. (a) Singlet state ($^1\text{A}$) PES calculated with different CAS sizes. (b) Singlet and triplet state ($^3\text{B}_1$) PESs calculated by the TCCSD(24,25) method.}
  \label{tcc_plots}
\end{figure*}

In Figure \ref{mrcc_plot}, we present the MR MkCCSD, MkCCSDT and DMRG(24,25) singlet  ($^1\text{A}$) and triplet state ($^3\text{B}_1$) TME PESs. TCCSD PESs are shown in Figure \ref{tcc_plots}.

As can be observed in Figure \ref{mrcc_plot}, the MR MkCCSD method gives wrong state ordering for all points along the twisting process except for one ($0^{\circ}$). Inclusion of triple excitations (MR MkCCSDT) in fact corrects this behavior. Nevertheless, nor the MR MkCCSDT method with CAS(2,2) model space is able to properly describe the PES of the singlet state with the apparent maximum at the torsional angle of $45^{\circ}$. There is actually an indication of an arising maximum on the MR MkCCSDT singlet PES close to $45^{\circ}$, however it is still too flat. Either enlargement of the model space as suggested by Pozun \textit{et al.} \cite{pozun_2013} or inclusion of higher (quadruple) excitations is probably necessary for the correct singlet state description. We have not pursued any of these possibilities, mainly due to considerably higher computational demands. Moreover, larger model spaces [in our case ideally CAS(6,6)] are not recommended for the Hilbert space multireference coupled cluster methods due to the so called proper residual problem \cite{lyakh_2012}.

Figure \ref{mrcc_plot} also depicts the DMRG(24,25) PESs. One can see the correct ordering of both spin states, however the singlet ($^1\text{A}$) PES also does not possess the right shape. The singlet state energy is correctly increasing when going from the torsional angle of $0^{\circ}$ to $45^{\circ}$, but does not sufficiently decrease for $45^{\circ}$ to $90^{\circ}$. Apparently, the missing dynamic correlation has an important effect on this part of the singlet PES, changing its shape qualitatively.

The TCC results from Figure \ref{tcc_plots} indicate that the TCCSD method is successful in recovering a major part of the missing dynamic correlation and thus properly describes the singlet PES. The effect of enlarging CAS is graphically depicted in Figure \ref{tcc_plot1} and numerically by comparison with the FCIQMC benchmark in Table \ref{tcc_tab}. One can observe that the value of the twisting energy barrier [$E(45^{\circ}) - E(0^{\circ})$] is decreasing with enlarging CAS and also improving towards the FCIQMC benchmark, eventually giving an excellent agreement for TCCSD(24,25) with the error of $-0.005$ kcal/mol.

\begin{table}[!ht]
 \begin{tabular}{lcc}
    \hline
    Method             & $E$ [kcal/mol] & $\Delta E$ [kcal/mol]      \\ \hline
    TCCSD(6,6)         & $2.527$ & $\phantom{-}1.395$         \\
    TCCSD(12,12)       & $1.577$ & $\phantom{-}0.445$         \\
    TCCSD(24,25) & \phantom{ }$1.127$ &  \phantom{ }$-0.005$ \\
    \hline
  \end{tabular}
  \caption{The TME singlet state ($^1\text{A}$) twisting energy barrier calculated by the TCCSD method with various CASs and the energy differences from the FCIQMC benchmark.}
  \label{tcc_tab}
\end{table}

Nevertheless, the shape of the singlet PES is only a part of the story.
In case of the spin state ordering, even the TCCSD method is not completely free of problems.
When calculating the high spin triplet component ($m_s = 1$), TCCSD(24,25) gives a wrong ordering of both spin states near the torsional angle of $45^{\circ}$ (see Figure \ref{tcc_plot2}). The reason for this behavior is obviously a fact that at $45^{\circ}$, both spin states very much differ in their character, triplet dominated by a single determinant, whereas singlet being strongly multireference with the two determinants (HOMO$^2$LUMO$^0$ and HOMO$^0$LUMO$^2$ \footnote{We use the standard notation of HOMO being the highest occupied molecular orbital and LUMO the lowest unoccupied molecular orbital.}) of practically equal weight. This is actually the worst case scenario for the TCC method, which even though we call a multireference CC, strictly speaking uses a single reference determinant and may be slightly biased in such ``degenerate'' situations. Taking into account that the TME singlet-triplet energy gap is really small ($0.2$ kcal/mol by FCIQMC), the aforementioned fact results in wrong state ordering.

To verify our assumptions, we have also calculated the low spin triplet component ($m_s = 0$). It is strongly multireference as well, since it can be qualitatively described by a combination of two determinants (HOMO$^{\alpha}$LUMO$^{\beta}$ and HOMO$^{\beta}$LUMO$^{\alpha}$) with equal weights. Our aim was to eliminate to some extent the bias towards one of the two equally important determinants by calculating the states of a similar character. Figure \ref{tcc_plot2} proves that such an approach gives correct spin state ordering.

Last but not least, Table \ref{tab_results} compares the singlet-triplet energy gaps for the torsional angles of $45^{\circ}$ and $90^{\circ}$ calculated by different methods with the DMC result of Pozun \textit{et al.} \cite{pozun_2013} and available experimental data. One can notice results of a similar quality for the MR MkCCSDT and TCCSD(24,25)$_{m_s = 0}$ methods, where the fact that the former method includes single, double, and triple excitations whereas the later only single and double excitations should be emphasized.

\begin{table}[!ht]
  \begin{tabular}{lcc}
    \hline
    \multirow{2}{*}{Method}       & \multicolumn{2}{c}{$\Delta E_\text{T-S}$ [eV]}  \\ \cline{2-3} 
                                  & \phantom{aa}$45^\circ$ & \phantom{aa}$90^\circ$ \\ \hline
    MkCCSD                        &           $-0.12$ &            $-0.03$  \\
    MkCCSDT                       & $\phantom{-}0.07$ & $\phantom{-}0.15$ \\
    DMRG(24,25)                   & $\phantom{-}0.05$ & $\phantom{-}0.13$ \\
    TCCSD(24,25)$_{\text{m}_\text{s}=1}$ &           $-0.09$ & $\phantom{-}0.04$ \\
    TCCSD(24,25)$_{\text{m}_\text{s}=0}$ & $\phantom{-}0.07$ & $\phantom{-}0.20$ \\
    FCIQMC                        & $\phantom{-}0.01$ & $\phantom{-}0.13$ \\
    best available               & $\phantom{-^a}0.02^a$ & $0.13 \pm 0.013^b$ \\ \hline
  \end{tabular}
  \caption{The TME singlet-triplet energy gaps corresponding to the torsional angles of $45^{\circ}$ and $90^{\circ}$ calculated by different methods. $^a$DMC result \cite{pozun_2013}, $^b$photo-electron spectroscopy result \cite{clifford_1998}.}
  \label{tab_results}
\end{table}

\section{Conclusions}
\label{section_conclusions}

We have presented the FCIQMC benchmark data for the twisting process of the TME diradical which give an excellent agreement with the previous DMC and available experimental results. Our computations verified that there is a maximum on PES of the ground singlet state ($^1\text{A}$) corresponding to the torsional angle of $45^{\circ}$. At this point, there is also the smallest vertical singlet-triplet energy gap of $0.01$ eV as provided by FCIQMC.

Against the FCIQMC benchmark data, we have critically assessed the accuracy of the MR MkCC and TCC methods. We have found out that the MR MkCCSD method is not able to correctly predict the ordering of both lowest lying spin states and that the MR MkCCSDT, though giving good values for the singlet-triplet energy gap, is, due to the small CAS(2,2) model space, not able to properly describe the shape of the multireference singlet PES. On the other hand, the TCCSD method describes the ground singlet state with an excellent accuracy, but for the correct ordering requires computation of the low-spin component of the triplet state ($^3\text{B}_1$).

Taking into account strengths and weaknesses of the employed MR CC approaches, we propose a combination of both of them, namely a multireference generalization of the tailored CC method. Such an approach, based on the Jeziorski-Monkhorst ansatz (Eq. \ref{jeziorski-monkhorst}), i.e. employing different sets of CC amplitudes for each reference determinant, and MR cluster analysis \cite{paldus-ccanalysis} of the MPS wave function, is currently being developed by some of us and will be a subject of the follow-up paper.

\section*{Acknowledgment}

We would like to thank J. Brabec for helpful discussions.
This work has been supported by the Czech Science Foundation (grant no. 16-12052S), Czech Ministry of Education, Youth and Sports (project no. LTAUSA17033),
the Hungarian-Czech Joint Research Project MTA/16/05,
the Hungarian  National  Research,  Development  and  Innovation  Office  (grant  no. K120569),
and  the  Hungarian  Quantum  Technology  National  Excellence  Program  (project no. 2017-1.2.1-NKP-2017-00001).
FCIQMC computations were carried out on the Salomon supercomputer in Ostrava, we would therefore like to acknowledge the support by the Czech Ministry of Education, Youth and Sports from the Large Infrastructures for Research, Experimental Development and Innovations project ``IT4Innovations National Supercomputing Center - LM2015070".

\bibliography{references}
\bibliographystyle{achemso}

\end{document}